\documentclass{article}
\usepackage[utf8]{inputenc}
\usepackage{bbm}
\usepackage{amsfonts}
\usepackage{amssymb}
\usepackage{amsmath}
\usepackage{amsthm}
\usepackage{enumitem}
\usepackage[dvipsnames]{xcolor}
\usepackage{graphicx}
\usepackage{comment}
\usepackage{tikz}
\newtheorem{theorem}{Theorem}[section]

\newtheorem{lemma}[theorem]{Lemma}
\newtheorem{proposition}[theorem]{Proposition}
\newtheorem{remark}[theorem]{Remark}

\newcommand\cvn{{\operatorname*{\longrightarrow}_{n\rightarrow\ii}}}
\newcommand{\norm}[1]{\left\| #1\right\|}
\newcommand{\parent}[1]{\left(#1\right)}
\newcommand{\set}[1]{\left\{ #1\right\}}
\newcommand{\av}[1]{\left| #1\right|}

\newcommand{\bra}[1]{\left( #1\right)}
\def\integral{\int_{\mathbb{R}^2}}
\newcommand{\ayoub}[1]{{\color{red} Ayoub: #1\\}}
\newcommand{\done}[1]{{\color{blue} Seen: #1\\}}
\newcommand{\real}[1]{\mathcal{R}e\left\langle#1\right\rangle}

\def\curl{\operatorname{curl}}
\def\div{\operatorname{div}}

\def\RR{\mathbb{R}}
\def\ZZ{\mathbb{Z}}
\def\CC{\mathbb{C}}
\def\NN{\mathbb{N}}

\newcommand\ii{{\infty}}

\def\dx{\mathrm{d}\mathbf{x}}
\def\x{\mathbf{x}}

\def\p{\mathbf{p}}
\def\A{\mathbf{A}}
\def\B{\mathbf{B}}
\def\bnull{\mathbf{0}}
\def\balpha{\boldsymbol{\alpha}}
\def\bsigma{\boldsymbol{\sigma}}
\def\u{\mathbf{u}}

\def\cE{\mathcal{E}}
\def\cA{\mathcal{A}}
\def\cS{\mathcal{S}}
\def\cZ{\mathcal{Z}}

\usepackage{subfiles}
\usepackage{hyperref}

\title{Stability of the one electron atom Schrödinger model with magnetic field in two dimensions}
\author{Ayoub Arraji$^{1}$, Saad Benjelloun$^2$,  Salma Lahbabi$^{1,3}$}
\date{
{
\small $^{1}$ College of Computing, Université Mohamed 6 Polytechnique, Benguerir, Morocco.\\
$^{2}$ Makhbar mathematical sciences research institute, Casablanca, Morocco.\\
$^{3}$ EMAMI, LRI, ENSEM, Université Hassan II, Casablanca, Morocco.\\
}
\vspace{0.4cm}
\today
}

\begin{document}
\maketitle

\begin{abstract}
We study  the stability of the  one  electron  atom Schrödinger model with self-generated magnetic field in two dimensions. The magnetic  energy is taken of the general form $K\int_{\RR^2} |B|^p$ and we study the stability of the model as a function of the power $p$ and the coupling constant $K$. We show that for $p>3/2$, the model is always stable, and  for  $p<3/2$, the model is always unstable. In the critical case $p=3/2$, there is a critical stability constant $K_c$, that we characterize in terms of zero modes of the Dirac-Weyl operator. The value of $K_c$ is approximated using analytic and numerical methods. 
\end{abstract}
	
\tableofcontents
	
\section{Introduction and main results}
 
One of the most important  triumphs of quantum mechanics is the explanation of the stability of matter, i.e., the fact that the energy of a given system is finite (stability of the first kind) $E_0 > -\infty$ ~\cite{Schrodinger1926}, and is asymptotically  bounded from below by a linear function of the number of particles (stability of the second kind) $E_0(N) \geq -C\,N$~\cite{Dyson}.

In the non relativistic Schrödinger model, when the interaction between particles is modeled with the Couloumb  potential $V(\x) = \frac{1}{|\x|}$, the uncertainty principle guarantees the stability  by making the kinetic energy increase for states concentrating around the nucleus. When the atom interacts with a magnetic field, stability only holds under a condition on the fine structure constant $\alpha$~\cite{FLL1, LL2,LLS}. The reason why there may be instability is the existence of the so-called zero modes of the Dirac-Weyl operator, which   are states with zero kinetic energy. Three dimensional zero modes  had first been exhibited in~\cite{LY3} and studied  subsequently in numerous works such as~\cite{adam1999zero,adam2000degeneracy,ES,dunne2008abelian} (see below for more details).


In the one electron atom case, a critical $K_{c,3}$ has been exhibited in~\cite{FLL1}, below which there is stability and above which there is instability. An upper and lower bound of $K_{c,3}$ are also given, which insure the stability of the model for the physically relevant constant.  The case of the one-electron-many-atoms, and one-nucleus-many-electrons systems were studied  in~\cite{LL2}, and the general molecule case was studied in~\cite{LLS}. In these cases,  sufficient conditions for stability are given. The model for a self-generated magnetic field has also been treated in the framework of the  Hartree Fock model~\cite{comelli2017hartree} and the reduced  Hartree Fock model~\cite{gontier_lahbabi}. In this last work, a characterisation of the critical constant $K_{c,3}^{\rm rHF}$ is given as a function of the number of electrons $N$. 

In~\cite{FLL1} and~\cite{gontier_lahbabi}, the critical constant is given as the optimum of a variational problem posed on the set of zero modes. In three dimensions, zero modes are rare~\cite{balinsky2001zero, elton2002local}, and they are very difficult to obtain. This  makes the estimation of the critical constant not accurate; for example, in~\cite{FLL1}, there is a factor $10$ between the upper and lower bound.   In two dimensions, zero modes are abundant as per the Aharonov-Casher theorem, a symbolic method described for the first time in~\cite{AC} and made rigorous in~\cite{erdHos2002pauli}. Our work aims to study the stability of matter in dimension two and benefits from the abundance of zero modes to find an accurate estimate of the critical constant $K_c$. We restrict ourselves to the one electron atom model, similar to the one studied in~\cite{FLL1}. 

\medskip
Let us first recall the results of~\cite{FLL1}. In three dimensions, the kinetic operator is the Pauli operator on $L^2(\RR^3,\CC^2)$
$$
\bra{\bsigma \cdot (\p+\A) }^2 = (\p+\A)^2+ \sigma\cdot \B,
$$
where $\p=-i\nabla$ is the momentum operator and  $\sigma=(\sigma_1,\sigma_2,\sigma_3)$ are the Pauli matrices
	\begin{align*}
		\sigma_1 = \begin{pmatrix}0 && 1 \\ 1 && 0\end{pmatrix}, && \sigma_2 = \begin{pmatrix} 0 && -i \\ i && 0\end{pmatrix}, && \sigma_3 = \begin{pmatrix}1 && 0 \\ 0 && -1\end{pmatrix}. 
	\end{align*}
For a magnetic field $\B\in L^2(\RR^3,\RR^3)$, it was shown in~\cite{FLL1} that there exists a unique  vector potential $\A\in L^6(\RR^3,\RR^3)$ satisfying 

\begin{equation*}	\curl \A=B\quad  \text{and}\quad \div \A=0. 
\end{equation*}
Assuming that the nucleus of charge $z$ is located at the origin $\x = 0$, the total energy of an electron in the state described by a spinor  $\psi = \begin{pmatrix}
\psi_1 \\ \psi_2
\end{pmatrix} \in H^1\bra{\mathbb{R}^3,\CC^2}$ is
\begin{equation*}
\cE_{K,3}(\psi, \A) :=  \int_{\mathbb{R}^3}|(\sigma\cdot\bra{\mathbf{p} + \mathbf{A}})\psi|^2-  z\int_{\RR^3}\frac{|\psi(\x)|^2}{|\mathbf{x}|}\dx +  K\int_{\RR^3} |B|^2 .
\end{equation*}
The first term is the kinetic energy, taking into account  the interaction of the magnetic field with the spin of the electron. The second term is the potential energy of the Couloumb force generated by the nucleus. The last term is the coupling with the magnetic energy, which represents in some sense the cost of creating the magnetic field. The coupling constant $K$ is equal to $(8\pi \alpha^2)^{-1}$, where $\alpha$ is the fine structure constant. In the system of units where $e=1$, $\hbar= 1$ and $m=1$, we have $\alpha=1/c\simeq 1/137$. Although $K$ has a fixed value, it is considered as a variable and the stability of the model is studied as a function of $K$. 
The instability is caused by the existence of zero modes of the Pauli operator, which are the normalized solutions $(\psi, \A)$  of the Dirac-Weyl equation 
$$\bsigma\cdot(\p + \A)\psi = 0.$$
Indeed, for a zero mode, the kinetic energy vanishes, and the remaining terms, i.e., the magnetic energy and the potential energy, scale as the inverse of a length. When the coupling constant $K$ is small enough, the total energy is negative and goes to $-\infty$ by the effect of scaling. This critical constant is then given as the optimum of a functional on the set of zero modes. It is shown in~\cite{FLL1} that for 
$$
K_{c,3}:= \sup\left\{\frac{z\int_{\mathbb{R}^3}\frac{|\psi|^2}{|\mathbf{x}|}\dx}{\int_{\mathbb{R}^3}|B|^{2}\dx},\; (\psi, \A) \text{ a zero mode }\right\},
$$
we have
\begin{itemize}
\item if $K>K_{c,3}$, then $\cE_{K,3}$ is bounded from below,
\item if $K<K_{c,3}$, then $\cE_{K,3}$ is not bounded from below. 
\end{itemize}
In three dimensions, a first example of zero modes was exhibited  in~\cite{LY3}. It is given by 
\begin{equation}\label{eq:3D-ZM}
\psi(\x)=
\frac{1 + i\sigma \cdot \x}{(1+|\x|^2)^{\frac32}} \phi_0,
\quad
\A(\x)=3 \, \parent{1+|\x|^2} 
\langle \psi, \sigma\psi \rangle,
\quad
\B(\x) = 12 \, \langle \psi, \sigma \psi \rangle,
\end{equation}
where $\phi_0$ is a normalizing spinor. Afterward, many other zero modes have been constructed and several methods were explored to produce them \cite{adam1999zero,adam2000degeneracy,ES,dunne2008abelian}. However, until now there is no general rule to characterize their existence, and they are still difficult to obtain because of their rarity~\cite{elton2002local,balinsky2001zero}, which  makes the computation of the critical constant $K_{c,3}$  delicate. For example, in~\cite{FLL1}, a bounding of $K_{c,3}$ is proved, with a factor $10$ between the upper and the lower bound. 
The lower bound obtained using the Loss-Yau zero mode~\eqref{eq:3D-ZM} is 
$K_{c,3} \geq \frac{1}{9\pi^3}z = 0.0035z$. The upper bound is found using different functional inequalities that ensure stability above the value 
$K_{c,3} \leq \frac{z}{8\pi} = 0.0397z$.
This result was improved by a factor of $3/4$ using a diamagnetic inequality for zero modes proved in~\cite{frank2022magnetic}. Combining techniques from~\cite{FLL1} and~\cite{frank2022magnetic}, we could improve the bound by a factor of $4/9$ (see Appendix~\ref{app:3d-case}). Hence, to the knowledge of the authors, the best known bounds are 
\begin{equation}\label{eq:K3}
    0.0035z \leq K_{c,3} \leq 0.0176z.
\end{equation}
 
\medskip
In this paper we are mainly interested in the one electron atom Schrödinger model with self generating magnetic field in two dimensions. This corresponds to an electron confined in the $(x_1, x_2)$-plane subject to the Coulomb potential created by a nucleus of charge $z$ and a magnetic potential $\B=B {\bf e_3}$ pointing in the $x_3$ direction and depending only on $x_1$ and $x_2$.
In this case the kinetic energy operator is the Pauli operator acting on $L^2(\RR^2,\CC^2)$
\begin{equation}\label{eqn:E_p}
(\sigma \cdot \bra{\p+\A})^2= \bra{\p+\A}^2 + B \, \sigma_3 ,
\end{equation}
where the vector potential $\A\in L^2_{\rm loc}(\RR^2,\RR^2)$ satisfies 
\begin{equation}\label{eq:A-B}
\curl \A=B\quad  \text{and}\quad \div \A=0.
\end{equation}

		
Here, we consider a general expression of the magnetic energy of the form $K\int_{\RR^2} \av{B}^p$, with $p=2$ being the physically relevant case, and $K$ being the coupling constant. We are interested in studying the model when $p$ and $K$ vary. 
For $1<p<2$, the set of admissible vector potentials $\A\in L_{\rm loc}^2(\RR^2,\RR^2)$ satisfying $\operatorname{curl}\A=B\in L^p(\RR^2)$ and $ \div\A=0$ is well characterized as (see Appendix~\ref{app:potential})
$$\mathcal{A}^p= \set{\A\in L^{\frac{2p}{2-p}}(\RR^2,\RR^2), \; \div\A=0}.
$$  
For $p\geq 2$ and $p=1$, we consider
$$
\cA^p=\set{\A\in L^2_{\rm loc}(\RR^2,\RR^2),\; \curl\A\in L^p(\RR^2),\; \div\A=0}.
$$
Note that for $p=1$, there might not be an $\A\in L^2_{\rm loc}(\RR^2,\RR^2)$ that generate $B\in L^1(\RR^2)$~\cite{erdHos2002pauli}. In order to accommodate measure valued magnetic potentials, an alternative definition of the Pauli operator is proposed in~\cite{erdHos2002pauli} using quadratic forms, but we do not treat this case here. 
		
		
The total energy of a spinor $\psi = \begin{pmatrix}
\psi_1 \\ \psi_2
\end{pmatrix} \in H^1\bra{\mathbb{R}^2,\CC^2}$ is then 
\begin{equation}\label{eq:energy}
\cE_K^{p}(\psi, \A) :=  \int_{\mathbb{R}^2}|(\mathbf{p} + \mathbf{A})\psi|^2+\int_{\RR^2}  B\langle \psi, \sigma_3 \psi \rangle-  z\int_{\RR^2}\frac{|\psi(\x)|^2}{|\mathbf{x}|}\dx +  K\int_{\RR^2} |B|^p .
\end{equation}
The first term is the kinetic energy, the second is the Zeeman term which represents the energy of the interaction of the magnetic field with the spin of the electron. Also, the first two terms add up to the Pauli kinetic energy $\integral\av{\sigma\cdot(\p + \A)\psi}^2 $. The third term is the potential energy from the Couloumb force generated by the nucleus of charge $z$ and located at the origin $\x=0$. The last term is the coupling with the magnetic energy. This energy is invariant under the following transformations: 
\begin{itemize}
\item gauge transformation $\A \rightarrow \A + \nabla \phi $ and $\psi \rightarrow e^{-i\phi}\psi$,
\item rotation of the plane by angle $\theta$.
\end{itemize}
The set of admissible states $(\psi, \A)$  is then  
\begin{equation}			\mathcal{S}^p := \left\{ \psi \in H^1(\mathbb{R}^2, \mathbb{C}^2), \integral |\psi|^2\dx = 1 \right\}\times \cA^p.
\end{equation}
The ground state energy is denoted by  \begin{equation}\label{eq:min-prob}
E_K^p:=\inf\set{ \cE_K^p(\psi,\A),\; (\psi,\A)\in\cS^p}. 
\end{equation}

Our first main result gives the stability (and instability) of the system,
depending on the values of $p$ and $K$.
    
\begin{theorem}\label{th:stability}
For all $p > 3/2$ and for all $K > 0$, we have
\begin{equation*}
E_K^p > -\infty. 
\end{equation*}
For $p = 3/2$, there exists a critical constant $0<K_c<+\ii$ such that 
\begin{itemize}
\item if $K> K_c$,  then $	E_K^{3/2} > -\infty$,
\item if $K< K_c $, then $ E_K^{3/2} = -\infty$.
\end{itemize}
For $1 \leq p < 3/2$ and  for all $K > 0$, we have 
\begin{equation*}
E_K^p  = -\infty.
\end{equation*}
\end{theorem}
Hence, in two dimensions, the stability is always insured for the physically relevant case $p = 2$  and the criticality appears for $p = 3/2$. 

\begin{remark}
If the Coulomb potential $-1/\av{\x}$ is replaced by a potential $V$ such that $V_-\in L^{1+\epsilon}(\RR^2)+L^\ii(\RR^2)$, then, using similar techniques, we can show that $\cE^p_K$ is bounded from below for any $p\geq 2$ and $K>0$. Besides, if $V_-\in L^r(\RR^2)$, then  $\cE^p_K$ is bounded from below for any $p> 1+\frac1r$.  Note that the Coulomb kernel in $d=2$ satisfies $V_-(x)=\log(\av{x})1_{\av{x}\leq 1}\in L^r(\RR^2)$ for any $r>1$, thus the model is stable for any $p>1$. 
\end{remark}

The following theorem gives the conditions for the existence of a ground state. 

\begin{theorem}\label{th:minimizer}
For $3/2<p<2$, and for $p=3/2$ and $K>K_c$, the minimization problem~\eqref{eq:min-prob} admits a minimizer.
    		\end{theorem}

We see that for $1<p<2$, whenever the energy is bounded from below, there exists a minimizer. For $p\geq 2$, although the system is stable, we cannot prove the existence of a minimizer, because of the lack of control of the vector potential $\A$ by the magnetic field $B$.

Our following result characterizes the critical value $K_c$ in term of two-dimensional zero modes, which are normalized solutions to the Dirac-Weyl equation
\begin{equation}
    	\sigma\cdot(\mathbf{p} + \mathbf{A})\psi = 0.
\end{equation} 
In~\cite{ES}, a geometric approach is proposed to derive three-dimension zero modes from a given zero modes in $\RR^2$. By this geometric construction, the two dimensional zero mode corresponding to the three-dimension zero mode~\eqref{eq:3D-ZM} from \cite{LY3}, is given by 
\begin{align}\label{eq:zm2D}
\psi(\mathbf{x}) = \frac{1}{\sqrt{\pi}}\left( \begin{array}{c} 0 \\ \frac{1}{1 + |\mathbf{x}|^2}\end{array}\right), \; 
\mathbf{A}(\mathbf{x}) = \left(\begin{array}{c}
\frac{-2x_2}{1 + |\mathbf{x}|^2}\\[3mm] \frac{2x_1}{1 + |\mathbf{x}|^2}\end{array} \right)\; \text{and}  \;
B(\mathbf{x}) = \frac{4}{(1 + |\mathbf{x}|^2)^2} .
\end{align}
In dimension two, zero modes are abundant as per Aharonov-Casher theorem. Hence, we could  derive a good numerical approximation of  the critical constant by variational methods (see Section~\ref{bounds}). We denote by 
\begin{equation}
		\mathcal{Z} := \left\{(\psi, \A)\in \mathcal{S}^{3/2}, \; \sigma\cdot(\p + \A)\psi = 0 \right\}
\end{equation}
the set of zero modes for the case $p = 3/2$. 
      
\begin{theorem}\label{th:Kc}
The critical value $K_c$ is given by 
\begin{equation}\label{eq:def-Kc}
K_c= \sup\left\{\frac{z\int_{\mathbb{R}^2}\frac{|\psi|^2}{|\mathbf{x}|}\dx}{\int_{\mathbb{R}^2}|B|^{3/2}\dx},\; (\psi, \A) \in \mathcal{Z}\right\}.
\end{equation}
    			The maximization problem~\eqref{eq:def-Kc} is attained for some zero mode $(\psi,\A) \in \mathcal{Z}$ and satisfies
    			\begin{equation}
    				0.13 z \leq K_c \leq  0.166 z.
    			\end{equation}
    	
    		\end{theorem}
      In dimension two, the estimation of the critical constant in significantly better than the three dimensional case.
    			
\begin{remark}
The  zero mode~\eqref{eq:3D-ZM} proposed in~\cite{LY3} is very particular as it has many symmetries. It has been shown to optimize $\inf \|A\|_6$~over the set of magnetic fields that admit zero modes~\cite{frank2022sharp} and it is conjectured to optimize $\inf \|B\|_{3/2}$ in the work of Frank and Loss~\cite{frank2022magnetic}. 
We note that the 2D version of this zero mode~\eqref{eq:zm2D} is not optimal for $K_c$. 
\end{remark}

In dimension two, a simple characterization of zero modes was given in~\cite{AC, erdHos2002pauli} when $B$ is integrable. The dimension of the kernel of the Dirac-Weyl operator is given in terms of the flux $\int_{\RR^2} B\dx$ and the solutions of the Dirac-Weyl equation are explicitly given in terms of the real generating potential function $\phi$ that satisfies the Laplace equation with $B$ as a source (see Section~\ref{sec:equation-EE-K-c}). 
This explicit construction allows us to derive, when $B$ is integrable, the Euler-Lagrange equation~\eqref{eq:euler_lagrange} for the problem defining the critical constant $K_c$. 
If $B$ is not integrable, there is no simple characterization of the solutions for the Dirac-Weyl equation as shown in~\cite{erdHos2002pauli}. 
    
\begin{theorem}\label{th:ELes}
Let 
\begin{equation}
\mathcal{Z}_1 := \left\{(\psi, \A)\in \mathcal{S}^{3/2}, \; \sigma\cdot(\p + \A)\psi = 0,\; B\in L^1(\RR^2) \right\},
\end{equation}
and 
\begin{equation}\label{eq:def-Kc1}
K_{c,1}= \sup\left\{\frac{z\int_{\mathbb{R}^2}\frac{|\psi|^2}{|\mathbf{x}|}\dx}{\int_{\mathbb{R}^2}|B|^{3/2}\dx},\; (\psi, \A) \in \mathcal{Z}_1\right\}.
\end{equation}
If the problem~\eqref{eq:def-Kc1} admits a maximizer, then the corresponding Euler-Lagrange equation  reads
\begin{equation}\label{eq:euler_lagrange}
\alpha(\psi, B) \, \frac{|\psi|^2}{|\x|} - \beta(\psi, B)\, |\psi|^2 + \gamma(\psi) \, \Delta(|B|^{1/2}) = 0,
\end{equation}
with 
\begin{equation}
\begin{cases}
\alpha(\psi, B) := 2 \left(\integral |B|^{3/2}\dx \right) \, \left(\integral |\psi|^2\dx \right) \\
\beta(\psi, B) := 2 \left(\integral |B|^{3/2}\dx\right) \, \left(\integral\frac{|\psi|^2}{|\x|}\dx \right)\\
\gamma(\psi) := \frac{3}{2} \left(\integral\frac{|\psi|^2}{|\x|}\dx\right) \, \left(\integral |\psi|^2\dx \right).
\end{cases}
\end{equation}
\end{theorem}
    		
The constant $K_{c,1}$ is obviously a lower bound of $K_c$ and if $K_c$ is attained for an $L^1$ magnetic field $B$, then $K_c=K_{c,1}$. 

\medskip
The remainder of the paper is organized as follows. In Section~\ref{sec:stability}, we detail the proof for the stability Theorem~\ref{th:stability} and in Section~\ref{sec:existence} we detail the one for the existence  Theorem~\ref{th:minimizer}. Section~\ref{sec:Kc} is devoted to the characterization of the critical constant $K_c$, in terms of zero modes, using numerical approximation and an improved diamagnetic inequality for zero modes in two dimensions. 
Throughout the paper, we use several functional inequalities, such as the Hydrogen uncertainty principle and Sobolev kind inequalities. We recall those in  Appendix~\ref{sec:inequalities}, and give a synthetic proof for some of them. 
In Appendix~\ref{app:potential}, we characterize the relationship between the magnetic field $B$ and the vector potential $\A$ in the two dimensional setting.  Finally, in appendix~\ref{app:3d-case}, we give the proof of our improved estimation \eqref{eq:K3} of the critical value in dimension three.
 
\medskip

\subsection*{Acknowledgment} The research leading to these results has received funding from OCP grant AS70 ``Towards phosphorene based materials and devices''. We  warmly thank David Gontier for stimulating discussions around this work.  S. Lahbabi
thanks the CEREMADE for hosting her during
the final writing of this article.

\section{Stability results: Proof of Theorem~\ref{th:stability}}\label{sec:stability}

The proof for Theorem \ref{th:stability} is divided into three steps. 
		
		\paragraph*{Step 1: Stability for $p>3/2$.}
		We will show that there exists a bounded from below and coercive real function $g_p$ such that for all $(\psi,\A)\in \cS^p$, we have
		$$
		\cE_K^p(\psi,\A)\geq g_p(\norm{\nabla \av{\psi}}_2). 
		$$
		Let $(\psi,\A)\in \cS^p$ and denote by  $X:=\norm{\nabla \av{\psi}}_2$. We first bound, from below, the Coulomb interaction term. By the Hydrogen uncertainty principle (see e.g.~\eqref{eq:hydrogen-inequality}) applied to $|\psi|$ with $N = 2$, we have 
		\begin{equation}\label{eq:potentia_control}
			-z\integral \frac{|\psi|^2}{|\x|}\dx \geq -2z\left(\integral |\nabla|\psi| \;|^2\dx\right)^{1/2}=-2zX. 
		\end{equation}
		Next, we bound from below the kinetic energy. We have  (recall~\eqref{eqn:E_p})
		\begin{equation}\label{eq:positive_energy}
			\integral |\sigma\cdot (\p + \A)\psi|^2=	\integral |(\p + \A)\psi|^2\dx +  \integral B\langle\psi, \sigma_3\psi\rangle \geq  0.
		\end{equation}
		Thus,
		\begin{equation*}
			\integral |(\p + \A)\psi|^2\dx \geq -  \integral B\langle\psi, \sigma_3\psi\rangle = -\int_{\RR^2}B\bra{|\psi_1|^2 - |\psi_2|^2}.
		\end{equation*}
		By interchanging the roles of $\psi_1$ and $\psi_2$, we deduce that
		\begin{equation}\label{eq:energy_fraction}
			\integral |(\p + \A)\psi|^2 \geq \left| \integral B\langle\psi, \sigma_3\psi\rangle \right|.
		\end{equation}
		Therefore, there exists $0 \leq \alpha \leq 1$ such that $\alpha\integral |(\p + \A)\psi|^2\dx = \left| \integral B\langle\psi, \sigma_3\psi\rangle  \right|$
		and 
		$$
		\integral |\sigma\cdot(\p + \A)\psi|^2\dx \geq (1 - \alpha)\integral |(\p + \A)\psi|^2 \geq (1-\alpha)X^2,
		$$
		where the last inequality is obtained using the diamagnetic inequality 
		$$\integral |\nabla|\psi| \;|^2 \dx \leq \integral |(\p + \A)\psi|^2\dx$$ (see e.g. \cite[Theorem 7.21]{LL}). We turn now to the magnetic energy. Using Hölder inequality, we have 
		$$\av{ \integral B\langle\psi, \sigma_3\psi\rangle}\leq  \norm{B}_p\norm{\langle \psi, \sigma_3\psi\rangle}_q ,  $$
		with
		$\frac{1}{p} + \frac{1}{q} = 1$. Besides, we have 
		$$\langle\psi, \sigma_3\psi\rangle \; = |\psi_1|^2 - |\psi_2|^2 \leq |\psi_1|^2+ |\psi_2|^2= |\psi|^2=:\rho. $$
		By Gagliardo-Nirenberg inequality (see e.g.~\eqref{eq:sobolev_2d}) there exists a universal constant $D_q=S_q^2$ such that 
		$$\norm{\rho}_q \leq \frac{1}{D_q^{1/q}}\left(\integral |\nabla|\psi| \;|^2 \dx\right)^{1/p}.$$
		Thus 
		$$
		\alpha X^2\leq \av{\integral B\langle\psi, \sigma_3\psi\rangle } \leq \frac{1}{D_q^{1/q}}\norm{B}_p  X^{2/p}
		$$
		and 
		$$
		\int_{\RR^2} \av{B}^p\geq  D_q^{p/q}\alpha^p X^{2p-2}.
		$$
		We finally find 
		\begin{equation}
			\cE^p_K(\psi, \A) \geq (1 - \alpha)X^2 + K D_q^{p/q}\alpha^pX^{2p - 2} - 2zX. 
		\end{equation}
We introduce the function $f_p(\alpha,x)=(1-\alpha)x^2 +C\alpha^p x^{2p-2}-2zx$. Optimizing over $\alpha\in[0,1]$, we find that 
$$-\alpha x^2 +C\alpha^p x^{2p-2}\geq - \left(C'x^{\frac{2}{p-1}}\right) 1_{x^{2(2-p)}\leq pC}  + \bra{Cx^{2(p-1)}-x^2} 1_{x^{2(2-p)}>pC},
$$
where $C'=\bra{p-1}\bra{p^{p/(p-1)}C^{1/(p-1)}}^{-1}$. Therefore 
$$
f_p(\alpha,x)\geq g_p(x):= \bra{x^2-C'x^{\frac{2}{p-1}}-2zx } 1_{x^{2(2-p)}\leq pC} + \bra{Cx^{2(p-1)}-2zx}1_{x^{2(2-p)}> pC}.
$$
For $p>3/2$ or for $p=3/2$ and $K>K_u=2z/D_q^{p/q}$, $g_p$ is bounded from below and coercive.  
		
		
	
  		\paragraph*{Step 2: Instability for $p<3/2$. }

			We will exhibit a sequence $(\psi_n,\A_n)\in \cS^p$ such that $\cE^p_K(\psi_n, \A_n) \longrightarrow -\infty$. Let  $(\psi,\A)$ be a zero mode. We have
			\begin{equation*}
				\cE^p_K(\psi, \A) =  - z\int_{\mathbb{R}^2}\frac{|\psi|^2}{|\mathbf{x}|}\dx +K\int_{\mathbb{R}^2}|B(\mathbf{x})|^p\dx.
			\end{equation*}
			We introduce the scaled sequences 
			\begin{align*}
				\psi_n(\mathbf{x}) &= n\psi(n\mathbf{x}),\quad 			\mathbf{A}_n(\mathbf{x}) = n\mathbf{A}(n\mathbf{x}),\quad \text{and}\quad 
				B_n(\mathbf{x}) = n^2 B(n\mathbf{ x}).
			\end{align*}
			We check that $(\psi_n,\A_n)$ is also a zero mode and $\curl \A_n=B_n$. Moreover, we have
			\begin{align*}
				\cE^p_K(\psi_n, \A_n) 
				&=  -n^2 z \int_{\mathbb{R}^2}\frac{|\psi(\mathbf{x})|^2}{|\mathbf{x}|}\dx+n^{2 p - 1 }  K\int_{\mathbb{R}^2}|B(\mathbf{x})|^p \,\dx.
			\end{align*}
			Hence, for $p < 3/2$ or for $p=3/2$ and $K$ smaller that $K_l(\psi,\A)=\frac{z\int_{\mathbb{R}^2}\frac{|\psi(\mathbf{x})|^2}{|\mathbf{x}|}\dx}{\int_{\mathbb{R}^2}|B(\mathbf{x})|^p}$, we have  $$\cE_K ^p(\psi_n, \A_n) \cvn-\infty.$$

  \paragraph*{Step 3: Criticality for $p=3/2$. }
We first note that the function $K\mapsto E_K$ (we discard the upper-script $3/2$ here) is the minimum of the functions $\cE_K$ that are linear and non-decreasing in $K$.  This shows that $K \mapsto E_K$ is concave and non-decreasing, hence it is continuous on its domain. We can define
$$K_c=\sup\set{K>0, E_K=-\ii}.$$
In the end of {\bf Step 1}, we have shown that if $K>K_u$, $E_K\neq -\ii$. Thus $K_c\leq K_u \neq  +\ii$. Besides, in the end of {\bf Step 2}, we showed that if  $K<K_l(\psi,\A)$, for any zero mode $(\psi,\A)$, then $E_K=-\ii$. Thus 

$$K_c\geq \sup\set{ K_l(\psi,\A),\;(\psi,\A)\in \cZ }> 0.$$ 
		
\section{Existence of ground states: Proof of Theorem~\ref{th:minimizer}}\label{sec:existence}

In this section, we prove that when the energy $\cE_K^p$ is bounded from below and $p<2$, then it admits a minimizer. We start by stating and proving a compactness result that will be useful in the rest of the paper. 
		
\begin{lemma}[Compactness result]\label{lem:compactness-result} Let $1<p<2$ and let $(\psi_n,\A_n)$ be a sequence in $\cS^p$ and let $B_n=\curl(\A_n)$. If, for some $C>0$, we have
$$\norm{\sigma\cdot \bra{\p+\A_n}\psi_n}_{L^2}+\norm{\av{\psi_n}}_{H^1}+\norm{B_n}_{L^p}<C,$$ 
then there exist $(\psi,\A)\in  \cS^p$ such that 
\begin{itemize}
\item $\A_n \rightharpoonup \A$ in $L^q(\RR^2,\RR^2)$, with $q=\frac{2p}{2-p}$,	
\item $\psi_n\rightharpoonup \psi$ in $H^1(\RR^2,\CC^2)$,
\item $\sigma\cdot \bra{\p+\A_n}\psi_n\rightharpoonup \sigma\cdot \bra{\p+\A}\psi$ in $L^2(\RR^2,\CC^2)$. 
\end{itemize}
As a consequence, we have
$$\norm{\psi}_2\leq 1 \quad \text{and}\quad\cE_K^p(\psi,\A)\leq \liminf  \cE_K^p(\psi_n,\A_n).$$
			
\end{lemma}

		\begin{proof}
			Since $(B_n)$ is bounded in $L^p$, so is $(\A_n)$ in $L^q$ (see Theorem~\ref{control_of_potential}). Thus, there exist weak limits, up to subsequences, $B\in L^p$ and $\A\in L^q$, which satisfy $\curl(\A)=B$ and $\div(\A)=0$ in the distributional sense and $\norm{B}_p\leq \liminf \norm{B_n}_p$.  		
			Besides, $(\av{\psi_n})$ is bounded in $H^1$. Thus, it admits a weak limit, up to a subsequence, $\widetilde{\psi}\in H^1$ which satisfies $\|\widetilde{\psi}\|_2\leq 1$. By the weak continuity of the potential energy~\cite[Theorem 11.4]{LL}, we have
			$$
			\int_{\RR^2}\frac{\av{\psi_n(\x)}^2}{\av{\x}}d\x \to \int_{\RR^2}\frac{|\widetilde{\psi}(\x)|^2}{\av{\x}}d\x.
			$$
   
We also have $(\av{\psi_n})$ (resp. $(\rho_n=\av{\psi_n}^2)$) that converges to $\widetilde{\psi}$ (resp. $\widetilde{\rho}=|\widetilde{\psi}|^2$), weakly in $L^r(\RR^2)$ and strongly in $L^r(\Omega)$ for any bounded $\Omega\subset\RR^2$ and any $2\leq r<\ii$ (resp. $1\leq r<\ii$), as $H^1(\Omega)$ is compactly embedded in $L^r(\Omega)$. 
			
			Finally, we note that for any $n\in \NN$, $\norm{\sigma\cdot\A_n \psi_n}_2\leq \norm{A_n}_q\norm{\rho_n}_{q'}^{1/2}$, with $2/q+1/q'=1$. Thus, the sequence $(\norm{\sigma\cdot\p \psi_n}_2=\norm{\nabla\psi_n}_2)$ is bounded, and $(\psi_n)$ admits a weak limit $\psi\in H^1$. Using a compactly supported test function, we obtain that  $\sigma\cdot \bra{\p+\A_n}\psi_n$ converges weakly in $L^2$ to $\sigma\cdot \bra{\p+\A}\psi$. The lower semi-continuity of the $L^2$-norm concludes the proof. 
		\end{proof}
		
		Using Lemma~\ref{lem:compactness-result}, we can now show the existence of minimizers. 
		\begin{proof}[Proof of  Theorem~\ref{th:minimizer}]
			
			Let $(\psi_n, \A_n)$ be a minimizing sequence for $\cE_K^p$. Let $K'<K$ for $p>3/2$ and $K_c< K'<K$ for $p=3/2$. We have
			\begin{align*}
				\cE_K^p(\psi_n,\A_n)=	\cE_{K'}^p(\psi_n,\A_n)+ (K-K')\int_{\RR^2}\av{B_n}^{p}\geq E_{K'}^p+(K-K')\int_{\RR^2}\av{B_n}^{p}.
			\end{align*} 
			It follows that $(B_n)$ is bounded in $L^p$. Besides, in the proof of Theorem~\ref{th:stability}, we have shown that there is a coercive function $g_p$ such that for $p>3/2$ and $\forall K\geq 0$ or for $p=3/2$ and $K> K_u$ 
			$$
			\cE_K^p(\psi_n,\A_n)\geq g_p(\norm{\nabla \av{\psi_n}}_2),
			$$ 
			Therefore $(\av{\psi_n})$ is bounded in $H^1$ in these cases. 
			As well, for $p=3/2$ and $K_c<K\leq K_u$, we have 
			\begin{align*}
				\cE_K^{3/2}(\psi_n,\A_n)&=\cE_{K_u}^{3/2}(\psi_n,\A_n)+(K-K_u)\int_{\RR^2}\av{B_n}^{3/2}\geq g_p(\norm{\nabla \av{\psi_n}}_2)-C.
			\end{align*}
			Therefore, $(\av{\psi_n})$ is bounded in $H^1$ in this case as well. 
			Therefore 
			\begin{equation*}
				\norm{\sigma \cdot (\p+\A_n)\psi}_{L^2}^2+	K\int_{\mathbb{R}^2} |B_n|^p\dx \leq C+ z \int_{\mathbb{R}^2} \frac{|\psi_n|^2}{|\mathbf{x}|}\dx \leq C+ 2z\norm{\nabla \av{\psi_n} }_2< C,
			\end{equation*}
			where we have used the Hydrogen atom uncertainty principle. Thus $(\psi_n, \A_n)$ satisfy the conditions for Lemma~\ref{lem:compactness-result} and we obtain the existence of $\psi\in H^1$ and $\A\in\cA^p$ such that 
			$$
			\norm{\psi}_2\leq 1 \quad \cE^p_K(\psi,\A)\leq E_K^p. $$
			
			We now show that $\psi\neq 0$ and can be normalized. Indeed, we know that the ground state energy of the non-magnetic problem is negative, thus there exists a normalized $\psi_0\in H^1$ such that $\cE_K^p(\psi_0,\bnull )<0$. It follows that $E_K^p < 0$ and  $\psi\neq 0$.  Let us  consider $\widetilde{\psi} = \frac{\psi}{ \norm{\psi}_2} $.  We have
			\begin{align*}
				0> E_K^p &\geq  \int_{\mathbb{R}^2}|\sigma\cdot(\mathbf{p} + \mathbf{A})\psi|^2\dx + K\int_{\mathbb{R}^2} |B|^p\dx -z\int_{\mathbb{R}^2} \frac{|\psi|^2}{|\mathbf{x}|} \dx\\
				&= \norm{\psi}_2^2\left(\int_{\mathbb{R}^2}|\sigma\cdot(\mathbf{p} + \mathbf{A})\widetilde{\psi}|^2\dx   -z\int_{\mathbb{R}^2} \frac{|\widetilde{\psi}|^2}{|\mathbf{x}|} \dx\right) +K\int_{\mathbb{R}^2} |B|^p\dx\\
    &\geq \int_{\mathbb{R}^2}|\sigma\cdot(\mathbf{p} + \mathbf{A})\widetilde{\psi}|^2\dx   -z\int_{\mathbb{R}^2} \frac{|\widetilde{\psi}|^2}{|\mathbf{x}|} \dx +
				K\int_{\mathbb{R}^2} |B|^p\dx =\cE_K^p(\widetilde{\psi},\A)
				\geq  E_K^p.
			\end{align*}
			So, we necessarily have $\norm{\psi}_2 =  1$, and 
			\begin{equation*}
				\cE_K^p(\psi, \A) = E_K^p.
			\end{equation*}
		\end{proof}

\section{Critical value $K_c$: Proof of Theorem~\ref{th:Kc}} \label{sec:Kc}
		
\subsection{Variational characterization of $K_c$}
		The proof of the variational characterization for the critical value of $K$ follows the same lines as in~\cite{FLL1}. We are interested in the critical case $p=3/2$, so, in this section,  we remove the upper-script $3/2$ when not necessary. We denote by 
		$$
		\widetilde{K}_c= \sup\left\{\frac{z\int_{\mathbb{R}^2}\frac{|\psi|^2}{|\mathbf{x}|}\dx}{\int_{\mathbb{R}^2}|B|^{3/2}\dx},\; (\psi, \A) \in \mathcal{Z}\right\}.
		$$
		In {\bf Step 3} of the proof of Theorem~\ref{th:stability}, we have shown that 
		$$
		\widetilde{K}_c\leq K_c.
		$$		
		Let now show that $\widetilde{K}_c \geq K_c$. 	The idea is to prove that for $K<K_c$, i.e., $\inf  \cE_K(\psi, \A) = -\infty$, then we can construct a zero mode $(\phi, \balpha)$ with magnetic potential $\beta=\curl(\balpha)$, such that 
		\begin{equation}\label{eq:zero_mode_inequality}
			K\int_{\mathbb{R}^2}|\beta(\mathbf{x})|^{3/2}\dx - z\int_{\mathbb{R}^2}\frac{|\phi(\x)|^2}{|\mathbf{x}|}\dx \leq 0,
		\end{equation}
		which would imply that $K_c \leq \widetilde{K}_c$. Assume that $ E_K = -\infty$ and let $(\psi_n, \A_n)\in \cS$ be a minimizing sequence. As the Coulomb term is the only negative term in the energy, we should have 	\begin{equation}\label{potential_energy_goes_to_infinity}
			z\integral \frac{|\psi_n(\mathbf{x})|^2}{|\mathbf{x}|}\dx \longrightarrow +\infty.
		\end{equation}
		It means that $\psi_n$ concentrates around the nucleus. We normalize this sequence by the scaling $\lambda_n = \norm{\nabla\psi_n}_2\to +\ii$ as follows
		\begin{align*}
			\psi_n(\mathbf{x}) &= \lambda_n \phi_n(\lambda_n \mathbf{x}),&
			\mathbf{A}_n(\mathbf{x}) &= \lambda_n\balpha_n(\lambda_n \mathbf{x}),&
			B_n(\mathbf{x}) &= \lambda_n^2 \beta_n(\lambda_n \mathbf{x}).
		\end{align*}
		These fields satisfy
  \begin{align*}
			\norm{\phi_n}_2 = 1 \quad \text{and  } \quad \norm{\nabla\phi_n}_2 = 1.
		\end{align*}
		Besides, for $n$  large enough, we have $\cE_K(\psi_n, \A_n) <0$, thus
  \begin{equation}\label{potential_energy_upper_bound}
			\int_{\mathbb{R}^2}|\sigma\cdot(\mathbf{p} + \mathbf{A_n})\psi_n|^2\dx + K \int_{\mathbb{R}^2}|B_n(\mathbf{x})|^{3/2}\dx \leq z\integral \frac{|\psi_n(\mathbf{x})|^2}{|\mathbf{x}|}\dx\leq 2z\lambda_n. 
		\end{equation}
  and 
		\begin{equation}\label{beta_phi_inequality}
			\lambda_n\int_{\mathbb{R}^2}|\sigma\cdot (\mathbf{p} + \mathbf{\balpha_n})\phi_n|^2\dx + K \int_{\mathbb{R}^2}|\beta_n|^{3/2}\dx \leq z\int_{\mathbb{R}^2}\frac{|\phi_n(\mathbf{x})|^2}{|\mathbf{x}|}\dx\leq 2z.
		\end{equation}
The sequence $(\phi_n,\balpha_n)$ satisfies the conditions in Lemma~\ref{lem:compactness-result}. Thus, there exists $\widetilde{\phi}\in H^1$ and $\balpha\in \cA$ such that $\|\widetilde{\phi}\|_2\leq 1$, and
		\begin{equation}
			\phi_n \overset{H^1}{\rightharpoonup} \widetilde{\phi}, \qquad
			\balpha_n \overset{L^6}{\rightharpoonup} \balpha \qquad \text{and} \qquad
			\sigma\cdot\bra{\p+\balpha_n}\phi_n\overset{L^{2}}{\rightharpoonup} \sigma\cdot\bra{\p+\balpha}\phi. 
		\end{equation}
  Besides, $\lambda_n\int_{\mathbb{R}^2}|\sigma\cdot (\mathbf{p} + \mathbf{\balpha_n})\phi_n|^2\dx$ is bounded, and since $\lambda_n \rightarrow +\infty$, then 
		\begin{equation*}
			\norm{\sigma\cdot (\mathbf{p} + \mathbf{\balpha})\phi}_2\leq \liminf	\norm{\sigma\cdot (\mathbf{p} + \mathbf{\balpha_n})\phi_n}_2= 0.
		\end{equation*}
		Thus, 
		\begin{equation}\label{zero_mode}
			\sigma\cdot(\mathbf{p} + \balpha)\widetilde{\phi} = 0.
		\end{equation}
		It remains to show  that $\widetilde{\phi} \neq 0$  and then normalize it. Indeed, we have
		\begin{align*}
			1 - \norm{\balpha_n}_6
   = \norm{\nabla\phi_n}_2 - \norm{\nabla\phi_n}_2\norm{\balpha_n}_6
   &\leq \norm{\nabla \phi_n}_2 - C\norm{\balpha_n\phi_n}_2\\
	&\leq  C	\norm{\sigma\cdot(\mathbf{p} + \balpha_n)\phi_n}_2 \to 0.
		\end{align*}
		We deduce that $(\norm{\balpha_n}_6)$ is bounded from below by a positive constant. 
		Thus 
		$$
		0< C\leq 	\norm{\balpha_n}_{6}\leq C \norm{\beta_n}_{3/2}\leq  C\int_{\mathbb{R}^2}\frac{|\phi_n(\mathbf{x})|^2}{|\mathbf{x}|}\dx\to C\int_{\mathbb{R}^2}\frac{|\widetilde{\phi}(\mathbf{x})|^2}{|\mathbf{x}|}\dx.
		$$
		Thus $\phi\neq 0$. 	We set $\phi = \frac{\widetilde{\phi}}{\norm{\widetilde{\phi}}_2}$. Hence, $(\phi, \balpha)\in\cZ$ and 
		by the lower semi-continuity of the $L^{3/2}$ norm, and the fact that $ \norm{\tilde{\phi}}_2 \leq 1$, we have
		\begin{align*}
			K \int_{\mathbb{R}^2}|\beta|^{3/2}\dx \leq  
   \liminf K \int_{\mathbb{R}^2}|\beta_n|^{3/2}
   &\leq C\lim\int_{\mathbb{R}^2}\frac{|\phi_n(\mathbf{x})|^2}{|\mathbf{x}|}\dx\\
   &=  C \int_{\mathbb{R}^2}\frac{|\widetilde{\phi}(\mathbf{x})|^2}{|\mathbf{x}|}\dx
	\leq z\int_{\mathbb{R}^2}\frac{|\phi(\mathbf{x})|^2}{|\mathbf{x}|}\dx.
		\end{align*}
		Therefore~\eqref{eq:zero_mode_inequality} is hence proved, thus $K \leq K_c$, which concludes the proof.

The proof that $K_c$ is attained for a certain zero mode follows the same line as the previous proof; the detailed proof can be read in~\cite{these:Ayoub}.

		\subsection{Bounding $K_c$}\label{bounds}
		
In this section, we bound the critical value $K_c$.
		
\paragraph*{Upper bound. }	

In order to find a good upper bound, we use an improved diamagnetic inequality for zero modes, similar to the three dimension inequality proved in~\cite{frank2022magnetic}.

\begin{lemma}\label{lem:Dia-ZM}
    For a zero mode $(\psi,\A)\in \cZ$, we have, almost everywhere, 
    \begin{equation}\label{eq:diamagnetic}
    |(\p + \A)\psi|^2 \geq 2\av{ \nabla|\psi|}^2.
\end{equation}
\end{lemma}
The proof of Lemma~\ref{lem:Dia-ZM} follows the same line as the one of~\cite[Lemma 3.1]{frank2022magnetic} for three dimensional fields. It can be read in~\cite{these:Ayoub}. 
We can now prove our upper bound.

\begin{proposition}\label{thm:2d-upper}
The critical constant has the upper bound
    \begin{equation}
        K_c \leq 0.166z.
    \end{equation}
\end{proposition}
\begin{proof}
    For a zero mode, we have
$$\integral |\bsigma\cdot(\p + \A)\psi|^2 = 0.$$
Using Lemma~\ref{lem:Dia-ZM} and the same techniques as before, we obtain
\begin{align*}
2\int_{\RR^2}\av{\nabla\av{\psi}}^2 \leq \integral |(\p + \A)\psi|^2 
\leq\norm{B}_{\frac32}\norm{\rho}_{L^3}\leq \frac{1}{S_2^{1/3}}\norm{B}_{\frac32}\bra{\int_{\RR^2}\av{\nabla\av{\psi}}^2}^{\frac23}.
\end{align*}
Thus, 
$$\integral \frac{|\psi|^2}{|\x|}  \leq 2\parent{\integral |\, \nabla|\psi|\,|^2  }^{1/2} \leq \frac{1}{\sqrt{2S_2}}\int_{\RR^2}\av{B}^{3/2}.
$$
Using the lower bound $S_2\geq 18.28$ proved in Theorem~\ref{th:Sobolev}, we obtain $K_c \leq 0.166z$.

\end{proof}

\paragraph*{Lower bound. }

To obtain a lower bound on $K_c$, we evaluate the functional $K_l$ with several zero modes. First, we test the historical zero mode~\eqref{eq:zm2D}
\begin{align*}
			\psi(\mathbf{x}) = \frac{1}{\sqrt{\pi}}\begin{pmatrix}0 \\\frac{1}{1 + |\mathbf{x}|^2} \end{pmatrix}, \, \mathbf{A}(\mathbf{x}) = \left(- \frac{2x_2}{1 + |\mathbf{x}|^2}, \frac{2x_1}{1 + |\mathbf{x}|^2} \right), \,
			B(\mathbf{x}) = \frac{4}{(1 + |\mathbf{x}|^2)^2},
\end{align*}
which is the 2D zero mode constructed from the 3D zero mode~\eqref{eq:3D-ZM} introduced in~\cite{LY3}, by the method from~\cite{ES}. It has been proved that it optimizes $\inf \|\A\|_6$, for fields $\A$ that generate zero modes~\cite{frank2022sharp}, and it is conjectured to give the optimal value of $\inf \|\mathbf{B}\|_{3/2}$ on the set of magnetic fields that generate zero modes~\cite{frank2022magnetic}. 

\begin{figure}
\centering
\includegraphics[scale = .6]{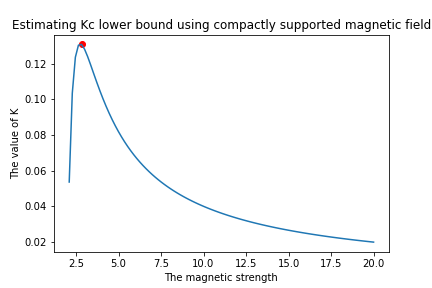}
\caption{The plot of the function $K(b)=K_l(\psi_b,\A_b)$. We observe that the best value is $K \approx 0.13z$ for the magnetic field strength $b \approx 2.82$.}
\label{fig:K-values}
\end{figure}

The evaluation of $K_l$ yields to the lower bound 
\begin{equation}
K_c  \geq K_l(\psi,\A)= \frac{1}{8}z= 0.125z.
\end{equation}
On the other hand, we consider the family of zero modes where the spinor is of the form
$$
\psi_{\alpha,\beta}(\x)=C\begin{pmatrix}0 \\\frac{1}{\bra{1 + |\mathbf{x}|^\beta}^\alpha} \end{pmatrix}.
$$
The best result for this family is obtained for $\alpha=2.77$ and $\beta=0.594 $ and gives the lower bound $K_c \geq 0.1308 $.
\\
We obtained a  similar lower bound  with magnetic field of the form $B_b(\x)=b{1}_{\av{\x}\leq 1}$  with $b>2$. The corresponding vector potential and spinor are given by     
$$\begin{cases}
A_b^1(\x) = -\frac{bx_2}{2}\bra{{1}_{\av{\x}\leq 1} +\frac{1}{x_1^2 + x_2^2}{1}_{\av{\x}> 1}}\\
A_b^2(\x) = \frac{bx_1}{2}\bra{{1}_{\av{\x}\leq 1}+\frac{1}{x_1^2 + x_2^2}{1}_{\av{\x}> 1}}
\end{cases}$$
and
$$ \begin{cases}
\psi_b^1(\x)=0\\
\psi_b^2(\x)=
e^{-\frac{b}{4}|\mathbf{x}|^2}1_{|\mathbf{x}| \leq 1}+ \left(\frac{1}{|\mathbf{x}|}\right)^{b/2}e^{-b/4} 1_{ |\mathbf{x}| \geq 1}.
\end{cases}$$
The value of $K_l$ can be explicitly calculated as a function of $b$ (see figure~\ref{fig:K-values}) and 
the best result is obtain for $b_0 \approx 2.82$. Thus 
$K_c \geq K_l(\psi_{b_0},\A_{b_0}) \approx 0.13z$. The relevant
code with symbolic calculations used to evaluate the different bounds can be found on Github\footnote{ \url{https://github.com/sa3dben/zero\_modes} }.

		\subsection{Euler-Lagrange equation: Proof of Theorem~\ref{th:ELes}}\label{sec:equation-EE-K-c}
		
		In dimension two, a simple characterization  of zero modes was given in~\cite{AC, erdHos2002pauli}. We use this explicit construction to derive the Euler-Lagrange equation~\eqref{eq:euler_lagrange}. 	Without loss of generality, we may assume that $\Phi= \frac{1}{2\pi}\int_{\RR^2} B>0$. If $ \Phi<1$, then  the corresponding Dirac equation does not have any non-zero solution. Otherwise ($\Phi\geq 1$), the solutions for the Dirac equation are explicitly given in terms of the "potential" $\phi$ satisfying 
$\textcolor{red}{-}\Delta \phi=B$, as 
 $$\psi=\bra{\begin{array}{c}0\\Pe^{-\phi} \end{array}},$$
where $P$ is a polynomial. 
Therefore 
$$
K_{c,1}=\sup\set{ K(\phi,P), \; \phi\in \mathcal{F},\; P \text{ polynomial such that } Pe^{-\phi}\in L^2(\RR^2)  }, 
$$	
where 
$$
K(\phi,P)=	\frac{z\int_{\mathbb{R}^2}\frac{|P e^{-\phi(x)}|^2}{|\mathbf{x}|}d\mathbf{x}}{\int_{\mathbb{R}^2}|\Delta \phi|^{3/2}\integral |P e^{-\phi}|^2\dx},
$$
and	the set of admissible potentials $\phi$ is
$$\mathcal{F} := \{\phi\in L^1_{\rm loc}(\RR^2), \; \Delta\phi \in L^{3/2}(\RR^2)\}. $$	
If $\Phi\notin\ZZ$, then the set of admissible polynomials is $\mathcal{P}_{[\Phi]-1}(X)$, the set of polynomials of degree less or equal to $[\Phi]-1$, while if $\Phi\in\ZZ$ then the set of admissible polynomials is $\mathcal{P}_{\Phi-1}(X)$ or $\mathcal{P}_{\Phi-2}(X)$. In order to derive the Euler-Lagrange equation, we consider the first order equation $$\delta \ln{K} = 0, $$ where $\delta$ is the functional derivative with respect to $\phi$. 
Denoting by 
\begin{equation*}
\psi := \begin{pmatrix}0 \\
Pe^{-\phi} \end{pmatrix}\quad \text{and} \quad 
B := -\Delta \phi,
\end{equation*} 
the optimality condition becomes 
\begin{equation}\label{eq:logarithm_K}
\frac{\delta \integral \frac{|\psi|^2}{|\x|}\dx}{\integral\frac{|\psi|^2}{|\x|}\dx} + \frac{\delta \integral |B|^{3/2}\dx}{\integral |B|^{3/2}\dx} - \frac{\delta \integral |\psi|^2\dx}{\integral |\psi|^2\dx} = 0.
\end{equation}
We have 
$$\delta |\psi|^2 = 2|\psi|^2 \delta\phi,$$
	thus 
	\begin{equation*}
		\delta \integral \frac{|\psi|^2}{|\x|}\dx = 2\frac{|\psi|^2}{|\x|}\delta \phi  \quad \text{and}\quad  \delta
		\integral |\psi|^2\dx = 2 |\psi|^2 \delta \phi.
	\end{equation*}
	The operator $\Delta$ is linear and symmetric, so
	$$\delta \integral |\Delta \phi |^{3/2}d\x = \frac32\Delta |\Delta\phi|^{1/2} \delta\phi= \frac32\Delta|B|^{1/2}\delta\phi. $$
	Using this in \eqref{eq:logarithm_K}, we obtain the desired result. 
 \medskip

The functional $K$ is symmetric under rotation, so we suspect the optimizer to be radially symmetric. In this case the Euler-Lagrange equation becomes
\begin{equation}
\balpha(\psi,B)\frac{|\psi|^2}{r} - \beta(\psi,B)|\psi|^2 + \gamma(\psi) \left(\frac{\partial^2|B|^{1/2}}{\partial r^2} + \frac{1}{r}\frac{\partial |B|^{1/2}}{\partial r}\right) = 0.
\end{equation}

\newpage
\appendix

\section{Some functional inequalities}\label{sec:inequalities}

Throughout the paper, we have used several functional inequalities, such as the Hydrogene uncertainty principle and Sobolev kind inequalities. In this section, we give a synthetic and novel proof of some useful functional inequalities in general dimension $N \geq 2$. A  Similar procedure  was proposed in~\cite{F_S}. 

\begin{theorem}\label{th:inequalities} For any $\psi\in H^1(\RR^N)$, we have 
				\begin{itemize}
					\item Heisenberg uncertainty principle: for $N\geq 1$, we have
					\begin{equation}
						\int_{\mathbb{R}^N}|\nabla\psi(\mathbf{\mathbf{x}})|^2\dx
						\int_{\mathbb{R}^N}|\mathbf{x}|^2|\psi(\mathbf{x})|^2\dx \geq \frac{N^2}{4}.
					\end{equation}
				\item Hydrogen uncertainty principle: for $N \geq 2$, we have
				\begin{equation}\label{eq:hydrogen-inequality}
					\int_{\mathbb{R}^N}|\nabla\psi(\mathbf{\mathbf{x}})|^2\dx\geq \frac{(N - 1)^2}{4}\left(\int_{\mathbb{R}^N}\frac{|\psi(\mathbf{x})|^2}{|\mathbf{x}|}\dx\right)^2.
				\end{equation}
		\item Hardy inequality: for $N \geq 3$, we have
			\begin{equation}
				\int_{\mathbb{R}^N}|\nabla\psi(\mathbf{\mathbf{x}})|^2\dx\geq  \frac{(N - 2)^2}{4}\int_{\mathbb{R}^N}\frac{|\psi(\mathbf{x})|^2}{|\mathbf{x}|^2}\dx.
			\end{equation}
		\item Linearized Sobolev inequality: for $N \geq 3$, we have
		\begin{equation}
			\int_{\mathbb{R}^N}|\nabla\psi(\mathbf{\mathbf{x}})|^2\dx\geq N(N - 2) \int_{\mathbb{R}^N} \frac{|\psi(\mathbf{x})|^2}{(1 + |\mathbf{x}|^2)^2}\dx.
		\end{equation}
				\end{itemize}
			\end{theorem}
			To prove Theorem~\ref{th:inequalities}, we use the following lemma. 
			
			\begin{lemma}\label{lem:inequalities}
				Let $\psi \in \mathcal{C}_c^\infty(\RR^N)$ and let $g$ be a real radial function, $g(\mathbf{x}
				)=G(\av{\mathbf{x}
				})$, with $G(r) = \mathcal{O}\left(\frac{1}{r^2}\right)$ and $rG'(r) = \mathcal{O}\left(\frac{1}{r^2}\right)$ in the neighborhood of $0$.  Then, we have the following inequality
				\begin{align*}
					&\left(\int_{\mathbb{R}^N}|\nabla\psi(\mathbf{\mathbf{x}})|^2\dx \right)
					\left(\int_{\mathbb{R}^N}|\mathbf{x}|^2 \, g(\mathbf{x})^2 \, |\psi(\mathbf{x})|^2\dx \right)\\
&\qquad\qquad\qquad      \qquad\qquad\qquad \geq \frac{1}{4}\left( \int_{\mathbb{R}^N} \left[ N\,g(\mathbf{x}) + |\mathbf{x}| \,  G'(\av{\mathbf{x}}) \, \right] \,  |\psi(\mathbf{x})|^2  \,  \dx\right)^2.
				\end{align*}
			\end{lemma}
   Before proving Lemma~\ref{lem:inequalities}, let us show how to use it to prove Theorem~\ref{th:inequalities}. 
   \begin{proof}[Proof of Theorem~\ref{th:inequalities}]
   We apply Lemma~\ref{lem:inequalities} for different functions $g$ as follows. 
			\begin{itemize}
	\item  Heisenberg uncertainty principle: we take $G(r) = 1$. 
	\item Hydrogen uncertainty principle: we take $G(r) = \frac{1}{r}$.
	\item Hardy inequality: we take $G(r) = \frac{1}{r^2}$.
	\item Linearized Sobolev inequality: we take $G(r) = \frac{1}{1 +r^2}$.
			\end{itemize}
   We extend the results to  $H^1(\RR^N)$ by  a density argument. 
		\end{proof}
 \begin{proof}[Proof of Lemma~\ref{lem:inequalities}]
Let $1 \leq k \leq N$, and let $\lambda$ be a real number.  We have
				\begin{equation*}
					F_k:=	\int_{\mathbb{R}^N} \, \av{ \bra{-i\partial_k - i\lambda \, x_k \, g(\mathbf{x}) } \, \psi(\mathbf{x})}^2\dx \geq 0. 
				\end{equation*}
				Besides, a simple calculation shows that 
				\begin{align*}
					F_k&=\int_{\mathbb{R}^N}|\partial_k\psi(\mathbf{x})|^2 \, \dx + \lambda^2\int_{\mathbb{R}^N}x_k^2 \, g(\mathbf{x})^2 \,  |\psi|^2\dx \\
     &\qquad\qquad\qquad \qquad\qquad\qquad -\lambda \int_{\mathbb{R}^N}\left(\frac{x_k^2}{|\mathbf{x}|}G'(\av{\mathbf{x}}) + g(\mathbf{x}) \right)|\psi(\mathbf{x})|^2\dx.
				\end{align*}
				We sum this inequality for all values of $1 \leq k \leq N$
				\begin{align*}
					&\sum_{k=1}^NF_k=	\int_{\mathbb{R}^N}|\nabla \psi(\mathbf{x})|^2\dx + \lambda^2\int_{\mathbb{R}^N}|\mathbf{x}|^2g(\mathbf{x})^2|\psi|^2\dx\\
     & \qquad\qquad\qquad \qquad\qquad\qquad - \lambda \int_{\mathbb{R}^N}\left(|\mathbf{x}|G'(\av{\mathbf{x}}) + Ng(\mathbf{x}) \right)|\psi(\mathbf{x})|^2\dx \geq 0.
				\end{align*}
				Optimizing over $\lambda\in\RR$, we obtain the desired inequality. 
			\end{proof}

We recall the folowing Sobolev type inequality, and give a lower bound on constants of interest for us. 
  			
			\begin{theorem}\label{th:Sobolev}
				For  $N\geq 2$ and $q \geq 1$, there is a constant $S_q^N > 0$ such that for any $\psi \in  H^1(\mathbb{R}^N, \mathbb{C}^2)$ with $\int_{\mathbb{R}^2}|\psi|^2 \dx = 1$, we have
				\begin{equation}\label{eq:sobolev_2d}
				\left(\int_{\mathbb{R}^{ N}} |\nabla \psi |^2\dx\right)^{\frac{N}{2}(q - 1)} \geq S_q^N \int_{\mathbb{R}^N}\rho^q \dx,
				\end{equation}
				with $\rho = |\psi|^2$. In particular, for $q=\frac{N+1}{N-1}$, we have 
    \begin{equation}\label{eq:sobolev_N}
    \left(\int_{\mathbb{R}^{ N}} |\nabla \psi |^2\dx\right)^{\frac{N}{N-1}} \geq S_N \int_{\mathbb{R}^N}\rho^{\frac{N+1}{N-1}} \dx,
    \end{equation}
    with $S_2\geq 18.28$ and $S_3\geq 24.05$.
			\end{theorem}
   
			\begin{proof}
	Inequality~\eqref{eq:sobolev_2d} is a direct consequence of~\cite[corollary 1.12]{F_S}. We will prove the particular case~\eqref{eq:sobolev_N}  and  give a lower bound on  $S_2$ and $S_3$. Let $e < 0 $ be
the ground state energy of $ -\Delta- {V(x)}$. In~\cite{lieb1991inequalities}, it is proved that
$$
\av{e}^{\frac12}\leq L_{N}\int_{\RR^N}\av{V}^\frac{N+1}{2}, 
$$
for some constants $L_{N}$. Thus, for a normalized $\psi\in H^1$, 
$$
-\int_{\RR^N}\av{\nabla\psi}^2+\int_{\RR^N}V\rho\leq -e=\av{e}.
$$
Thus,
$$\int_{\RR^N}|\nabla \psi|^2 \geq \int_{\RR^N} V\rho - L_{ N}^2 \left(\int_{\RR^N}  \av{V}^{\frac{N+1}{2}}\right)^2.$$
Taking  a potential of the form $V = C\rho^{\frac{2}{N-1}}$, we obtain 
$$\int_{\RR^N} |\nabla \psi|^2 \geq C\int_{\RR^N} \rho^{\frac{N+1}{N-1}} - C^{N+1}L_{ N}^2  \left(\int_{\RR^N} \rho^{\frac{N+1}{N-1}}\right)^2.$$
Optimizing over $C$ gives
$$
\bra{\int_{\RR^N} |\nabla \psi|^2}^{\frac{N}{N-1}} \geq 
\frac{N^{\frac{N}{N-1}}  }{\bra{1+N}^{\frac{N+1}{N-1}}L_{ N}^{\frac{2}{N-1}} } \int_{\RR^N} \rho^{\frac{N+1}{N-1}}. 
$$
Using the values in~\cite{lieb1991inequalities} $L_2 \approx 0.09$  and $L_3 \textcolor{red}{= } 0.0135$, we obtain 
$$
S_2\geq 18.28 \quad \text{and}\quad S_3\geq 24.05. 
$$
\end{proof}

\section{2D Magnetic vector potential}\label{app:potential}

			In this section, we characterize the vector potentials $\mathbf{A}$ corresponding to a magnetic field $B$.
			\begin{theorem}\label{control_of_potential}
				For a magnetic field $B \in L^p(\mathbb{R}^2)$, with $1 < p < 2$, there is a unique solution $\mathbf{A} \in L^{\frac{2p}{2 - p}}(\mathbb{R}^2, \mathbb{R}^2)$  to the equation
				$$\operatorname{curl}(\mathbf{A}) = B,$$ 
				satisfying $\div(\mathbf{A}) = 0$. This field is given by
				\begin{equation}\label{eq:defA}
					\mathbf{A}(\mathbf{x}) = \int_{\mathbb{R}^2} \mathbf{G}(\mathbf{x} - \mathbf{y})B(\mathbf{y})d\mathbf{y},
				\end{equation}
				where
				\begin{equation*}
					G(\mathbf{x}) = \frac{1}{2\pi}\bra{\frac{-x_2}{|\mathbf{x}|^2} \, , \frac{x_1}{|\mathbf{x}|^2}}.
				\end{equation*}
				Moreover, there is a constant $S$ independent of $\mathbf{A}$ and $B$ such that
				\begin{equation}\label{magnetic_inequality}
					\norm{\mathbf{A}}_q \leq S\norm{B}_p.
				\end{equation}
			\end{theorem}
			
			\begin{proof}
				We prove that 
				\begin{equation*}
					\mathbf{A}(\mathbf{x}) := \int_{\mathbb{R}^2} \mathbf{G}(\mathbf{x} - \mathbf{y})B(\mathbf{y})d\mathbf{y} \in L^{\frac{2p}{p - 2}}(\RR^2,\RR^2). 
				\end{equation*}
				Let  $\phi\in L^q(\RR^2)$ be a test function,   with $q=2p/(3p-2)$ being the Hölder conjugate of $\frac{2p}{2-p}$. We have 				
				$$\integral |\mathbf{A}(\x)\phi(\x)|\dx \leq \integral \frac{|B(\mathbf{y})|}{|\mathbf{x} - \mathbf{y}|}\av{ \phi(\x)}d\mathbf{y}\dx.$$
				Besides, $
					\frac{1}{p} + \frac{1}{q} + \frac{1}{2} = 2.
				$
				Thus, by the  Hardy-Litllewood-Sobolev inequality (see e.g. Theorem 4.3 \cite{LL}), there exists a universal constant $S>0$ such that  				
				$$\integral |\mathbf{A}(\x)\phi(\x)|\dx \leq S\norm{B}_p\norm{\phi}_q.$$
				 Therefore, $\mathbf{A}$ is indeed in $L^\frac{2p}{2 - p}$ and $$ 	\norm{\mathbf{A}}_{\frac{2p}{2 - p}} \leq S\norm{B}_p.$$ 
    It remains to prove that  $\mathbf{A}$ is unique. Let $\mathbf{A},\mathbf{A}'\in L^{\frac{2p}{2 - p}}(\mathbb{R}^2, \mathbb{R}^2)$ be two vector potentials satisfying
				\begin{equation*}
						\operatorname{curl}(\mathbf{A}) =  \operatorname{curl}(\mathbf{A}') = B, \quad \text{and}\quad 
						\operatorname{div}(\mathbf{A}) =  \operatorname{div}(\mathbf{A}') = 0.
				\end{equation*}
Hence, for $\mathbf{U}= \mathbf{A}-\mathbf{A'}$, 
				\begin{equation}
					\label{holomorph_condition}
						\operatorname{curl}(\mathbf{U}) = 0, \quad \text{and}\quad
						\operatorname{div}(\mathbf{U}) = 0.
				\end{equation}
It follows that $U_1$ and $U_2$ are harmonic functions. Since $U_1, U_2 \in L^{\frac{2p}{2-p}}$, then they vanish.
			\end{proof}


\section{Improved estimation for the three dimensional critical constant}\label{app:3d-case}

\def\B{\mathbf{B}}
The problem of criticality in~\cite{FLL1}  is formulated in term of the charge number $z$, and the critical value $z_c$ is given as
\begin{equation}
    z_c = \inf \left\{\frac{1}{8\pi\alpha^2}\frac{\int_{\RR^3}|\B(\x)|^2\dx}{\int_{\RR^3}\frac{|\psi|^2}{|\x|} \dx} \right\},
\end{equation}
with the infimum taken over all the normalized zero modes.

\begin{theorem}\label{th:improvement-3D}
    The critical charge number $z_c$ is bounded below by the numerical value $40,475$.
\end{theorem}
This value is an improvement over the value $17,900$ given in~\cite{FLL1} and the value $25,025$, given in~\cite[Theorem 1.5]{frank2022magnetic}. We still get a large factor of approximately $5$ between the upper bound $208,000$ and the lower bound $40,475$.

The proof of Theorem~\ref{th:improvement-3D} follows the same steps as its two dimensional counterpart in Proposition~\ref{thm:2d-upper}. Indeed, we find that for any zero mode, we have 
$$
\bra{\frac{3}{2}}^2S_3\int_{\RR^3}\frac{\av{\psi}^2}{\av{\x}}\leq \int_{\RR^3}\av{\B}^2.
$$
Thus 
$$
z_c\geq \frac{1}{8\pi\alpha^2}\bra{\frac{3}{2}}^2S_3=40475.
$$

\bibliographystyle{siam}
\bibliography{results,reviews,zero-modes} 
			
\end{document}